%% file: main.tex
\documentclass[10pt,conference]{IEEEtran}
\IEEEoverridecommandlockouts
\usepackage{cite}
\usepackage{amsmath,amssymb,amsfonts,bm}
\usepackage{algorithmic}
\usepackage{xcolor}
\usepackage{xcolor}
\usepackage{graphicx}
\usepackage{multirow}

\usepackage[linesnumbered,ruled,noend]{algorithm2e}

\SetAlgoCaptionSeparator{:}
\SetNlSty{texttt}{}{:}
\usepackage[T1]{fontenc}
\usepackage[font=small,labelfont=bf,tableposition=top]{caption}

\DeclareCaptionLabelFormat{andtable}{#1~#2  \&  \tablename~\thetable}

\SetAlgoCaptionSeparator{:}
\SetNlSty{texttt}{}{:}

\newcommand{\eg}{\emph{e.g.}\xspace}
\newcommand{\ie}{\emph{i.e.}\xspace}

\newtheorem{definition}{Definition}[section]

\def\BibTeX{{\rm B\kern-.05em{\sc i\kern-.025em b}\kern-.08em
    T\kern-.1667em\lower.7ex\hbox{E}\kern-.125emX}}
\begin{document}

\title{Model Synthesis for Communication Traces of System-on-Chip Designs\\
}

 \author{\IEEEauthorblockN{Hao Zheng, Md Rubel Ahmed}
 \IEEEauthorblockA{
 U of South Florida, Tampa, FL\\
 \{haozheng,mdrubelahmed\}@usf.edu}
 \and
 \IEEEauthorblockN{Parijat Mukherjee}
 \IEEEauthorblockA{Intel, Hillsboro, OR \\parijat.mukherjee@intel.com }
 \and
 \IEEEauthorblockN{Mahesh C. Ketkar}
 \IEEEauthorblockA{Intel, Folsom, CA \\
 mahesh.c.ketkar@intel.com}
 \and
 \IEEEauthorblockN{Jin Yang}
 \IEEEauthorblockA{Intel, Hillsboro, OR\\
  jin.yang@intel.com}
  }

\maketitle

\begin{abstract}

Concise and abstract models of system-level behaviors are invaluable in design analysis, testing, and validation.
In this paper, we consider the problem of inferring models from communication traces of system-on-chip~(SoC) designs.  The traces capture communications among different blocks of a SoC design in terms of messages exchanged.  The extracted models characterize the system-level communication protocols governing how blocks exchange messages, and coordinate with each other to realize various system functions.
In this paper, the above problem is formulated as a constraint satisfaction problem, which is then fed to a SMT solver.  The solutions returned by the SMT solver are used to extract the models that accept the input traces.
In the experiments, we demonstrate the proposed approach with traces collected from a transaction-level simulation model of a multicore SoC design and traces of a more detailed multicore SoC design developed in GEM5 environment. 





\end{abstract}



\input{src-introduction}

\input{src-literature.tex}
\input{src-background.tex}
\input{src-method.tex}

\input{src-result.tex}

\input{conclusion.tex}

\noindent{{\bf Acknowledgment}
The research presented in this paper was partially supported by gifts from the Intel Corporation, and a grant from Cyber Florida.





\bibliographystyle{unsrt}

\input{main.bbl}

\end{document}

%% file: src-introduction.tex
\section{Introduction}
\label{motivation}
Modern system-on-chip (SoC) designs integrate a large number of functional blocks procured from various sources.  At runtime, these blocks communicate and coordinate with each other through intricate system-level protocols to implement various sophisticated functions.  SoC executions are highly concurrent where a large number of system transactions following those protocols are often executed simultaneously.
Experiences have shown that communications among different blocks are the major source of various design and runtime errors. 
In order to thoroughly verify the communication behavior of an SoC, well-defined and comprehensive protocol specifications are essential.
However, in practice, such specifications are usually not available,  ambiguous, incomplete, or even contain errors. 
They often become outdated and disconnected from the design implementation as the design progresses. 

Many methods and approaches, \eg~\cite{Yang:2006,Liu:2013,Mrowca:2019:LTS:3316781.3317847,natasa2020}, have been proposed to mine patterns or models from traces of various forms.  They are inadequate to handle the SoC communication traces considered in this work.  The existing methods extract patterns of events from a trace if those events show strong temporal dependencies.  
However, in communication traces as a result from concurrent executions of a large number of system transactions, events showing strong temporal dependencies may not be related according to the true dependencies in the ground truth specifications. 
In the next section, an example is used to elaborate this challenge.  
As a result, existing mining methods often generate a large number of patterns, many of which are not meaningful.  
Moreover, they may not be able to extract a model that includes all real valid patterns. 
Therefore, the extracted models can be large, not understandable, or even misleading.

To address the above challenge, this paper describes a method that can automatically infer reduced and concise models from communication traces of SoC designs obtained from simulation or emulation.  
These traces considered in this paper are sequences of messages exchanged among design blocks.  
Given an input trace, the model extraction is formulated as constraint satisfaction problem, which is then fed to an SMT solver.  
A model is constructed from solutions returned by the SMT solver.
The extracted model characterizes the underlying system-level communication protocols that design blocks follow to generate the input trace. 

The main \textbf{contribution} of this work is, to our best knowledge, the first method that can automatically and efficiently synthesize concise abstract models from system communication traces for SoC designs.  
What distinguishes this work from previous ones is that it targets traces generated from concurrently executing multiple system transactions, and aims to infer system-level protocols that govern the executions of those system transactions.  
As a result, it avoids finding patterns in the extracted models without true dependencies by incorporating readily available design information, thus leading to reduced and more understandable models. 
It is efficiently, and scalable to very long traces.



%% file: src-literature.tex
\smallskip
\noindent{\bf Related Work.~}
Specification mining aims to extract patterns from various artifacts. 
A model based approach \textit{Synoptic} \cite{Beschastnikh:2011:LEI:2025113.2025151} mines invariants from logs of sequential execution traces where concurrency is recorded in partial order. It then generates an FSM that satisfies the mined invariants. 
\emph{Perracotta}~\cite{Yang:2006} is another software execution log analysis tool that mines temporal API rules.  It describes a chaining technique that can be used to find long sequential patterns.
The work \textit{BaySpec} \cite{Mrowca:2019:LTS:3316781.3317847} extracts LTL formulas from Bayesian networks trained with software traces. It requires traces to be clearly partitioned with respect to different functions.
For mining hardware traces, the approaches presented in~\cite{Li2010DAC,Hertz:2013:tcad,Danese:2015:vlsi-soc,Danese:2015:date,Danese:2017:dac} mine assertions from either gate-level representations~\cite{Li2010DAC}, or RTL models~\cite{Chang:2010:aspdac,Hertz:2013:tcad,Danese:2015:vlsi-soc,Danese:2015:date,Danese:2017:dac}.
The work in~\cite{Liu:2013} describes an assertion mining approach using episode mining from the simulation traces of transaction level models.
Those approaches often lack support for finding longer patterns, thus not able to find complex communication patterns involving multiple components.

Our work shares some similarities to the methods on software model synthesis.  The aim of model synthesis is to identify a model from system execution traces such that the resulting model can accept the input traces.
In~\cite{Heule:2013}, the deterministic finite automata inference based on the original evidence-driven state merging method~\cite{Lang:1998:EDSM} is extended and formulated as a graph coloring problem, which is then solved by a Boolean satisfiability solver.  Recently, \emph{Trace2Model} is introduced in~\cite{natasa2020} that learns non-deterministic finite automata (NFA) models from software execution traces using C bounded model checking technique. Similar work can also be found in~\cite{Ulyantsev:2011}
The above approaches do not consider the concurrent nature of communication traces of SoC designs, and they rely on temporal dependencies discovered from traces to identify models. However, temporal dependencies are often not same as the true dependencies that our work aims to find.  

%% file: src-background.tex
\section{Background}
\label{sec:background}



\begin{figure}[tb]
\begin{center}
\begin{tabular}{cc}
\begin{minipage}{1.8in}
\footnotesize
\begin{verbatim}
1 (cpu0:cache:rd_req)
2 (cache:cpu0:rd_resp)
3 (cpu1:cache:rd_req)
4 (cache:cpu1:rd_resp)
5 (cache:mem:rd_req)
6 (mem:cache:rd_resp)
\end{verbatim}
\end{minipage}
& 
\begin{minipage}{1.in}
\includegraphics[height=1.1in]{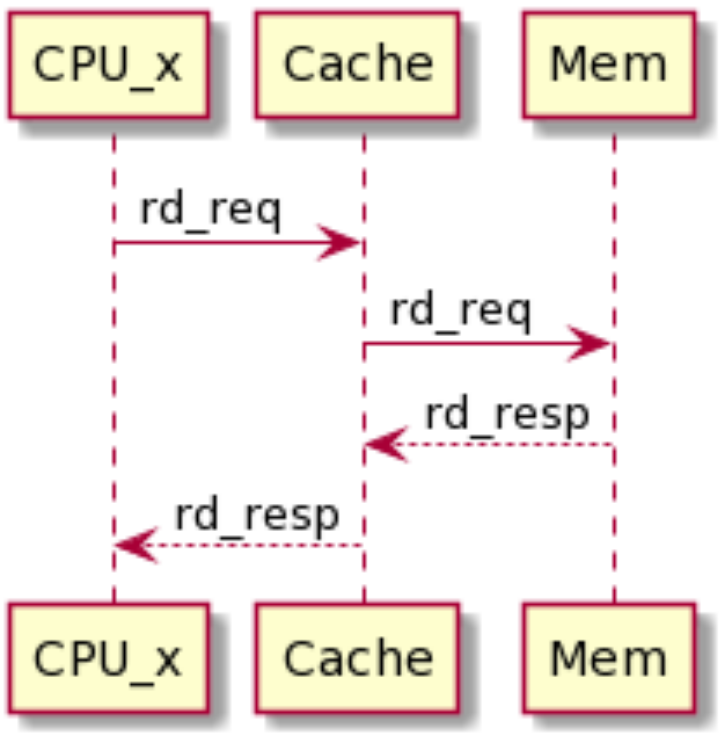}
\end{minipage}
\\
(a) & (b)
\end{tabular}
\vspace{-1pt}
\caption{CPU downstream read flows. (a) Definitions of messages, (b) Messages sequence diagram for the flows.  This diagram is parameterized with $x$ which can be $0$ or $1$.}
\label{fig:flow-ex}
\end{center}
\vspace{-15pt}
\end{figure}

System-level protocols are often represented as \emph{message flows} in SoC architecture documents.
Fig.~\ref{fig:flow-ex} shows a very simple yet representative example of message flows for a multicore SoC design.  This example specifies memory read operations for two CPUs via a shared cache.  Message flows specify temporal relations for a set of messages.  As shown in Fig.~\ref{fig:flow-ex}(a), each message is a triple $({\tt src: dest: cmd})$ where the ${\tt src}$ denotes the originating component of the message while ${\tt dest}$ denotes the receiving component of the message.  Field ${\tt cmd}$ denotes the operation to be performed at ${\tt dest}$.  
For example, message ${\tt (CPU\_0: Cache: rd\_req)}$ is the read request from ${\tt CPU\_0}$ to ${\tt Cache}$.  
Each message flow is associated with an unique \emph{start} message to indicate initiation of an instance of such flow, and with one or multiple different \emph{end} messages to indicate its completion. 
A  flow may contain multiple branches describing different paths a system can execute such a flow. For example, the flow in Figure~\ref{fig:flow-ex}(b) has two branches specifying read operations in cases of cache hit or miss.

During the execution of an SoC design, instances of flows are executed concurrently. 
When a flow instance is executed along one of its paths, messages on that path are exchanged with runtime information, \eg memory addresses.  Typically, multiple instances of different flows that are executed concurrently are captured in the traces. 

\begin{definition}
An SoC execution \textbf{trace} $\rho$ is 
    $\rho = (\varepsilon_0, \varepsilon_1, \ldots, \varepsilon_n)$
where $\varepsilon_i = \{m_{i,0},\ldots, m_{i,k}\}$ is a set of messages observed at time $i$, and $m_{i,j} $ is an message instance of some flow instance active at time $i$, for every $m_{i,j} \in \varepsilon_i$.
\end{definition}
To simplify the presentation, hereafter flows~(messages) and flow instances~(message instances) are used interchangeably if their meanings are clear in the context.  

An example trace from executing the flows in Figure~\ref{fig:flow-ex} is 
\begin{equation}
\label{ex:trace}
\left(\{1, 3\}, 1, 2, 5,  1, 5, 6, 2, 4, 6, 2\right)
\end{equation}
where the numbers in the trace are the message indices as shown in the Fig.~\ref{fig:flow-ex}(a). 
The idea of message sets helps to represent the outcomes of these concurrently executing flows. Note that the ordering of the messages in the same set of a trace is unknown.  Given two messages $m_i$ and $m_j$ and a trace $\rho$, we define $m_i <_\rho m_j$, denoting that $m_i$ occurs before $m_j$ in $\rho$, if $m_i \in \varepsilon_i$, $m_j \in \varepsilon_j$, and $i<j$.  For simplification, a set of a single message $\{m\}$ is written as $m$.  Moreover, a sequence of one message $(m)$ is also written as $m$.

Given a trace as shown above, we aim to infer a model that characterizes the underlying message sequence protocols as shown in Fig.~\ref{fig:flow-ex} to explain the generation of the trace.  We use finite state automata to represent the extracted model.


\begin{definition}
A finite state automaton (FSA) is a tuple $\mathcal{M} = (Q, q_0, \Sigma, F, \Delta)$  where $Q$ is a finite set of states, $q_0 \in Q$ the initial state, $\Sigma$ a finite set of symbols, $F \subseteq Q$ the set of accepting states, and $\Delta : Q \times \Sigma \to Q$ the transition relation.
\end{definition}

In our method, each symbol in $\Sigma$ denotes an unique message found in the input trace, and $F = \{q_0\}$.  
Given a FSA $\mathcal{M}$, a flow execution scenario is defined as a set $\mathcal{X} = \{(\mathcal{M}^i,q^i_j)~|~ 1 \leq i\leq n\}$ where $\mathcal{M}^i$ is the $i$th instance of $\mathcal{M}$, and $q^i_j$ is a current state of $\mathcal{M}^i$. 
Suppose $\rho$ is an input trace and $\mathcal{M}$ is the model inferred from $\rho$.
Initially, all $\mathcal{M}^i$ in $\mathcal{X}$ are in $q_0$.  
Then, for every message $m$  in $\rho$ from the beginning,  we can find an $(\mathcal{M}^i,q^i_j) \in \mathcal{X}$ such that $\Delta(q^i_{j}, m, q^i_{j+1})$ holds for some state $q^i_{j+1} \in Q$.
After accepting $m$, we get a new execution scenario $\mathcal{X}'$ with $(\mathcal{M}^i,q^i_j)$ being replaced with $(\mathcal{M}^i,q^i_{j+1})$.
At the end of the trace $\rho$, the corresponding execution scenario is the same as the initial execution scenario.  
Multiple instances of the model $\mathcal{M}$ in an execution scenario reflect the fact that the trace is the result from concurrent execution of multiple message flows when a SoC design runs. 



\begin{figure}
    \centering
    \includegraphics[width=3.4in]{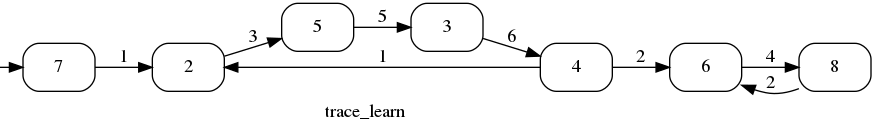}
    \caption{FSA model produced using the method in~\cite{natasa2020}}
    \label{fig:trace2model-output}
\end{figure}

The challenge of inferring models from communication traces as shown above is that such traces are results from executing many message flows concurrently, thus correlating messages correctly is very difficult.  
Note that $(1,5,6, 2)$ and $(3,5,6,4)$ are two message flows specified in Fig.~\ref{fig:flow-ex}. 
Consider the trace $(1,3,5,6,1,3,5,6,2,4,2,4)$ resulting from executing the above message flows two times each in an interleaved manner.  
In this trace, messages $1$ and $3$ show strong temporal dependency, but they are actually unrelated. 
Existing methods fail to extract true message flows because they mostly rely on temporal dependencies to find interesting patterns.  
For example, using $\mathit{Perracotta}$ in~\cite{Yang:2006} we can extract a sequential pattern $(1, 3, 5, 6)$, which is highly confusing. Fig.~\ref{fig:trace2model-output} shows the model produced using the method in~\cite{natasa2020}.  
Even though this model fits the trace perfectly, it contains message sequences that do not make sense with respect to the ground truth flows. 

Note that the message flows in Fig.~\ref{fig:flow-ex} specifies causality relations among messages in addition to their temporal relations.  For example, when message $\tt(CPU\_0:Cache:rd\_req)$ occurs, it causes either $\tt(Cache:CPU\_0:rd\_resp)$ or $\tt(Cache:Mem:rd\_req)$ to happen.  Therefore, we define  \emph{structural causality} below based on the observation: \emph{any message in an SoC execution trace is an output of a component in reaction to a previous input message.}  
\begin{definition}
\label{causal}
Message $m_j$ is {\bf causal} to  $m_i$, denoted as $\mathit{causal}(m_i, m_j)$, if $m_i{\tt .src} = m_j{\tt .dest}.$
\end{definition}

The above causality is referred to as \emph{structural} against the functional causality in the flow specifications.  
In our method, message sequences represented by the FSA model extracted from a trace are required to satisfy the structural causality relation for every two consecutive messages.  With this requirement, sequence $(1, 3, 2, 4)$ would not be extracted as a model of the above trace.  Applying the structural causality relation during the model extraction process is straightforward as it only depends on basic structural information of a SoC design, which can be readily captured in messages. 

%% file: src-method.tex
\section{Model Synthesis}
\label{sec:flowminer}


Algorithm~\ref{algo:flowminer} shows the outline of our method.  It takes as an input a trace of messages $\rho$, and produces a FSA model $\mathcal{M}$ such that executing $\mathcal{M}$ in a certain way can lead to $\rho$ to be re-produced.  This method involves three major steps, collecting messages and building a causality graph~(lines 3-4), generating constraint problem~$P$ from the causality graph~(line~5), and extracting a model from $P$~(lines 6-7). 
These steps are explained in the following sections. 

\setlength{\textfloatsep}{0pt}
\IncMargin{.5em}
\begin{algorithm}[tb]
\caption{\textbf{AutoModel}}
\label{algo:flowminer}
\textbf{Input:} A trace $\rho$\\
\textbf{Output:} {a FSA $\mathcal{M}$}\\
\SetAlgoNoLine
Extract unique messages in $\rho$ into $M$\;
Build the causality graph $G$ from $M$\;
Generate $P$ the set of consistency constraints from $G$\;
Get a solution $sol$ of $P$ using a SMT solver\;
Derive a FSA $\mathcal{M}$ from $sol$\;
\end{algorithm}
\DecMargin{.5em}

\subsection{Building Causality Graph}
\label{sec:cg}

Given an input trace $\rho$, it is scanned to collect all unique messages.
A message is unique if at least one of its three attributes is different from all other messages already collected.
Recall that each message flow is initiated with a start message, and completed with an end message. 
During the message collection process, start and end messages are also identified.  Start and end messages can be identified from a trace as follows.  A message $m$ is a start message if 
\begin{equation}
\label{eq:start-m}
m.{\tt src} \not= m'.{\tt dest} \mbox{ for all } {m'} < {m} \mbox{ in }\rho.
\end{equation}
A message is an end message if 
\begin{equation}
\label{eq:end-m}
    m.{\tt dest} \not= m'.{\tt src} \mbox{ for all } m < m'  \mbox{ in } \rho. 
\end{equation}
Scanning the trace from the beginning, we can find all start messages by checking condition~(\ref{eq:start-m}). Similarly, all end messages can be found by scanning the trace from the end and checking condition~(\ref{eq:end-m}). 

Let all the collected messages be $M$, start messages $Start \subset M$, and end messages $End \subset M$. 
Next, we construct a causality graph from those messages, which captures all possible structural causalities among every pair of messages.  A causality graph $G$ is a directed graph.  It has a set of root nodes, each of which is labeled with a start message in $Start$, and a set of terminal nodes, each of which is labeled with a end message in $End$.  The other nodes are labeled with with messages that are not start or end messages.  

Once the causality graph is constructed, the trace is scanned again to find the supports for nodes and edges.
The support of a node is the number of instances of the labeled message in the trace. The edge support is the number of co-existences of messages at the head and tail nodes of that edge in the trace.

\begin{figure}
    \centering
    \begin{tabular}{cc}
        \begin{minipage}{1.6in}
        \[
        \begin{array}{lcl}
         M & = & \{1,2, 3,4,5,6\}\\
         Start & =& \{1, 3\}\\
         End &=& \{2,4\}
        \end{array}
        \]
        \end{minipage}
         &
         \begin{minipage}{1.6in}
             \includegraphics[width=1.6in]{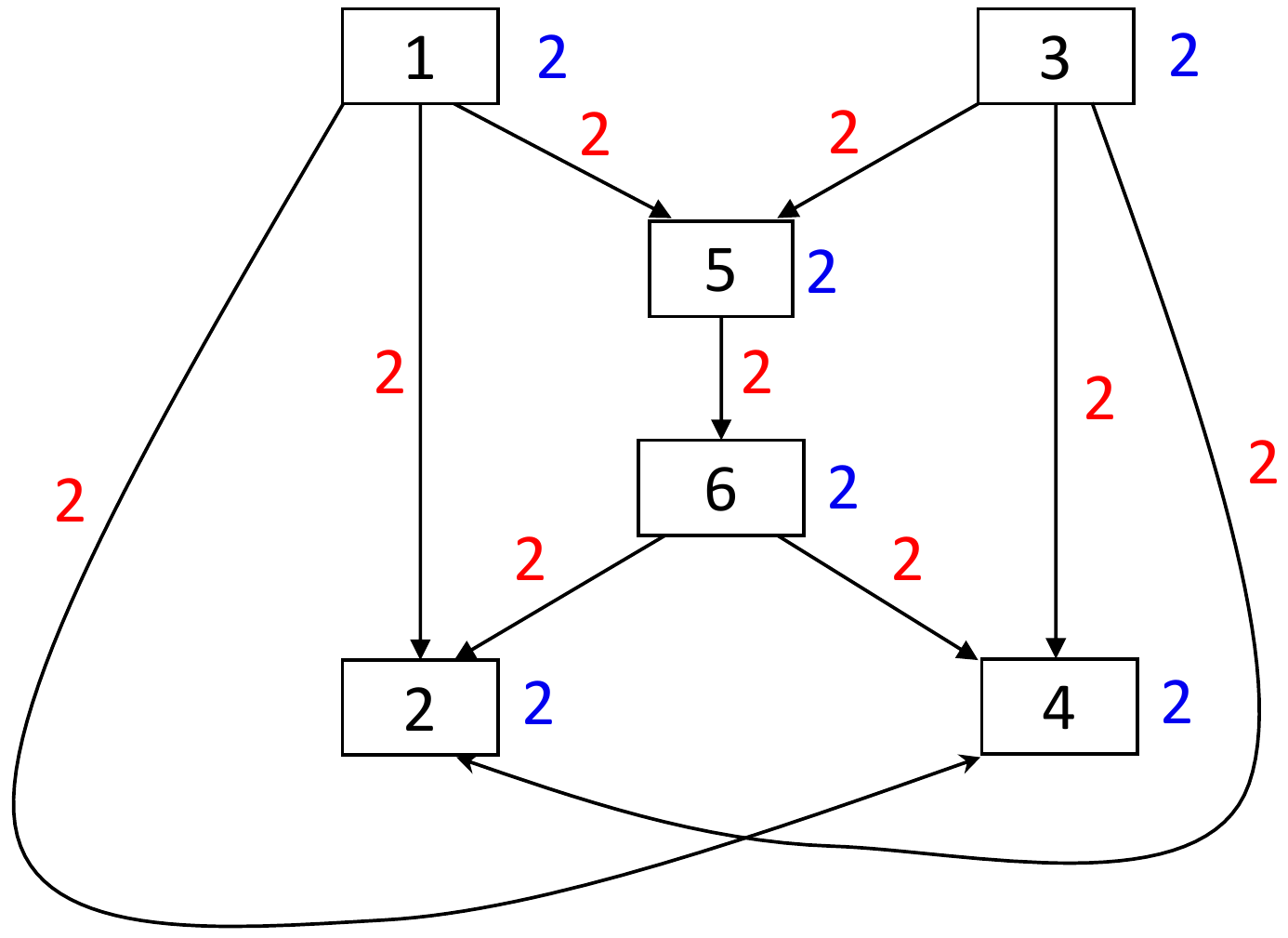}
        \end{minipage}
        \\
         (a) & (b)
    \end{tabular}
    \caption{Messages and causality graph constructed from a trace.}
    \label{fig:cg-ex}
\end{figure}

Consider the trace $\rho = (1,3,5,6,1,3,5,6,2,4,2,4)$ as an examples. The collected messages and the causality graph for this trace is shown in Figure~\ref{fig:cg-ex}.  In the causality graph, numbers in blue are node supports, and numbers in red are edge supports.  For example, we can find two instances of $(1,5)$ in the above trace, therefore, the edge from node 1 to node 5 in the causality graph is labeled with $2$.

\subsection{Generating and Solving Consistency Constraints}
\label{sec:constr-solve}

The constructed causality contains potential models for the input trace, however, it also contains a lot of inconsistencies. 
Consider node~1 and all its outgoing edges. The node support is $2$, and the total support of its outgoing edges is $6$.  This inconsistencies is due to the ambiguities when finding supports for edges. In this example, since we do not know what the other messages should be correlated with message~1, we consider all possibilities by finding supports for all binary sequences starting with message~1, \ie, $(1, 2)$, $(1,4)$, and $(1,5)$.  As a result, message~1 is counted more times than it should be. 
   
The goal of this step is to find a set of edge supports such that they are consistent with node supports. 
First, let $x\to y$ denote an edge from node $x$ to $y$, and $sup(\cdot)$ be the support of either a node or an edge.  
Additionally, let $c(n\to n')$ be a variable about the support of $n\to n'$. 
From the causality graph, we derive the following constraints.  

\begin{enumerate}
    \item For each node $n$, and all its outgoing edges $n\to n'$, create a constraint
        \[sup(n) = \sum_{\mbox{all } n\to n'} c(n\to n') \]
    \item For each node $n'$ and all its incoming edges $n \to n'$, create a constraint
            \[sup(n') = \sum_{\mbox{all } n\to n'} c(n\to n') \]
    \item For each edge $n\to n'$, create a constraint
            \[0 \leq c(n\to n') \leq sup(n\to n') \]
\end{enumerate}
The first two constraints require consistencies between every node and its incoming and outgoing edges.  The third constraint reflects the fact that it is unknown about the exact edge supports when scanning the trace for instances of an edge except it can be no more than the supports directly obtained from the trace.

\subsection{Deriving Model}  
\label{sec:model-gen}

\begin{figure}
    \centering
    \begin{tabular}{cc}
    \begin{minipage}{1.5in}
    \includegraphics[width=1.2in]{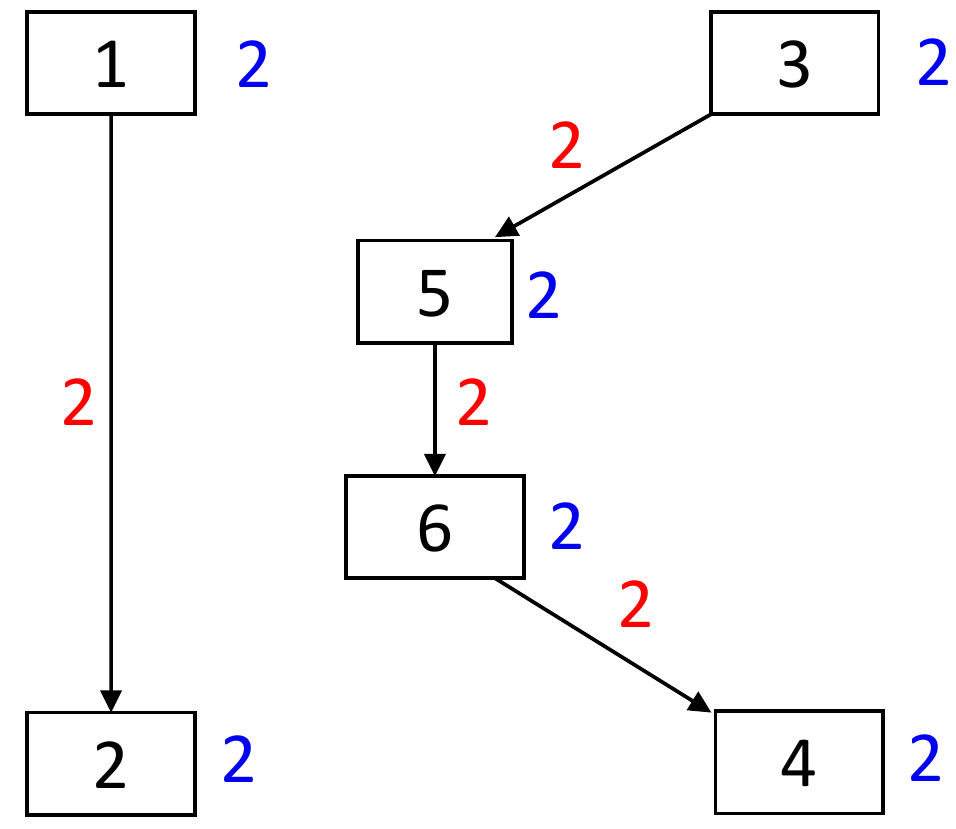}
    \end{minipage}
    &     
    \begin{minipage}{1.5in}
    \includegraphics[width=1.1in]{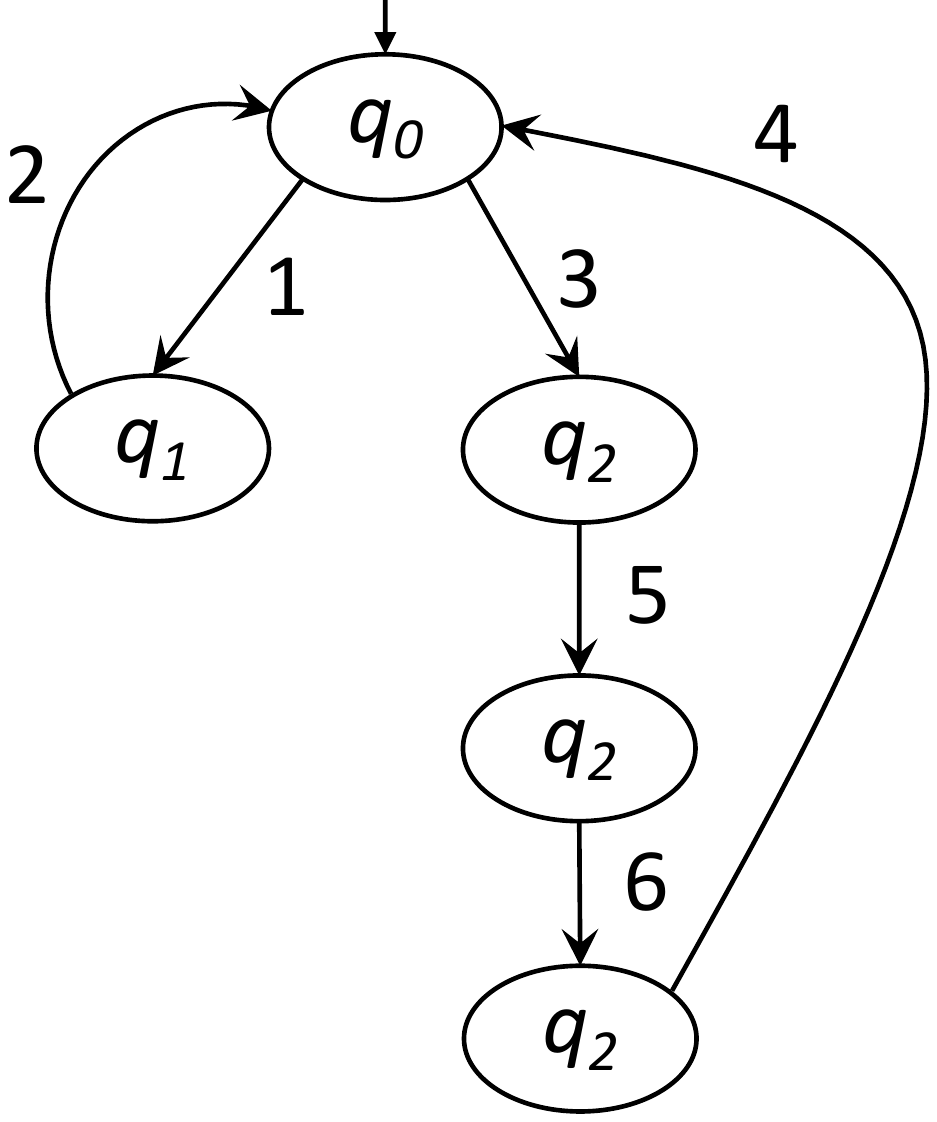}
     \end{minipage}
\\
      (a)   & (b) 
    \end{tabular}
    \caption{(a) The modified causality graph showing a consistent solution, and (b) the corresponding FSA model.}
    \label{fig:cg-sol-1}
\end{figure}

After the constraint problem $P$ is generated, it is fed into a constraint solver to find a solution.  A solution $\mathit{sol}$ of $P$ is a set of edges in the causality graph $G$ such that their supports are consistent with the node supports,
\[
\{(n\to m)~|~(n\to m) \mbox{ in } G, \mbox{ and } \mathit{c}(n\to m) > 0 \mbox{ in } \mathit{sol}\}.
\]   
The solution can be visualized as a modified causality graph with the edge supports as returned from the solver.  An example solution for the causality graph in Fig.~\ref{fig:cg-ex} is shown in Fig.~\ref{fig:cg-sol-1}(a) and the corresponding FSA model is shown in Fig.~\ref{fig:cg-sol-1}(b).
Note that in Fig.~\ref{fig:cg-sol-1}(a), edges from Fig.~\ref{fig:cg-ex} with zero-supports in the solution are removed. 

For a set of constraints $P$ generated in the previous step, a large number of consistent solutions can be generated.  Finding the minimal solution is NP-hard.  
Therefore, the goal of this step is to generate a reduced, not necessarily minimal, solution efficiently. 
In our method, we query the solver to return a set of solutions $S= \{sol~|~sol \models P\}$. 
Then, for each solution $\mathit{sol} \in S$, and for each edge $(n\to m)$ with non-zero support in $\mathit{sol}$, a new constraint is generated where its support is set to $0$.  
After adding this constraint into the solver, if the solver becomes unjustifiable, $\mathit{sol}$ is return as the candidate model. 
Otherwise, the above step repeats for the reduced solution $\mathit{sol}'$.
At the end, from the set of reduced solutions, we select the one with the smallest number of edges with non-zero support, and return it to the user.  
The model extraction method is shown in Algorithm~\ref{algo:model-extract}.
The model in Fig.~\ref{fig:cg-sol-1} shows an example model extracted using Algorithm~\ref{algo:model-extract}. 
Note that sequence $(1, 3)$ or $(2, 4)$ is not included in the model. 

After a reduced solution is returned as described above, a FSA can be constructed from it.
This step is straightforward, and we skip its explanation. 

\IncMargin{1.5em}
\begin{algorithm}[tb]
  \SetKwData{Left}{left}
  \SetKwData{Up}{up}
  \SetKwInOut{Input}{input}
  \SetKwInOut{Output}{output}
\caption{\textbf{ModelExtract}}
\label{algo:model-extract}

\Indm\Indmm
  \Input{Constraints $P$}
  \Input{Size limit of solutions $sz$}
  \Output{A reduced solution $\mathit{sol}$}
\Indp\Indpp
  \BlankLine
  $C = \emptyset$\;
  Find $S = \{\mathit{sol}~|~\mathit{sol} \models P\}$ of size $sz$\;
  \ForEach{$sol \in S$}{
    $sol' := \mbox{ReduceModel}(P, sol)$\;
    $C := C \cup \{ sol'\}$\;
  }
  Let $sol \in C$ with the minimal size\;
  return $sol$\;
\end{algorithm}
\DecMargin{1.5em}

\IncMargin{1.5em}
\begin{algorithm}[tb]
  \SetKwData{Left}{left}
  \SetKwData{Up}{up}
  \SetKwInOut{Input}{input}
  \SetKwInOut{Output}{output}
\caption{\textbf{ReduceModel}}
\label{algo:reduce-model}

\Indm\Indmm
  \Input{Constraints $P$}
  \Input{A solution $\mathit{sol}$}
  \Output{A reduced solution $\mathit{sol}'$}
\Indp\Indpp
  \BlankLine
    Get an edge $(n\to m)$ in $\mathit{sol}$\; 
    $P' := P \cup c(n\to m) =0$\;
    \If{$P'$ is unjustifiable}{\Return $\mathit{sol}$}
    Get a new solution $\mathit{sol}' \models P'$\;
    \Return ReduceModel($P'$, $\mathit{sol}'$)\;
\end{algorithm}
\DecMargin{1.5em}

%% file: src-result.tex
\section{Experimental Results}
\label{sec:results}

We collect ten message flows which are abstracted from system-level communication protocols used in real industry SoC designs.  Each flow consists of a number of branches, resulting in a total number of $64$ message sequences to specify various system communication scenarios such as cache-coherent memory accesses,  upstream read/write, power management, etc.  Although greatly simplified, these message flows capture essential communicating behaviors among typical components in a SoC design, including CPUs, caches, interconnect, memory controller, and peripheral devices, etc.  

We implement a transaction-level simulation model where components in the model communicate with each other according to the protocols specified by these message flows.  During simulation, each master component randomly initiates a message flow instance by generating a message, and sends it to the next component as specified. 
Since the message payloads are not considered in the message flows, they are not generated during the simulation, and each message only carries a command and the address of the destination component. In order to mimic the concurrent nature of modern SoC designs, the interconnect component of our simulation model is modeled as a switch network.  This component model allows messages from a source to go to all other destinations as specified. In case when messages arriving from multiple sources simultaneously, they are interleaved using an internal arbiter before being sent out. All components run concurrently without global synchronization. 

We simulate the model multiple times, collecting two sets of traces of different lengths.  The first set of traces are generated when the model is restricted to only allow CPUs to initiate cache coherent downstream read flows while all other components are configured to only react to incoming messages. This set of traces are referred to as {\sf small}.  The second set of traces are generated with all components in the model are enabled to initiate message flows.  This set of traces are referred to as {\sf large}.  For each configuration, we simulate the model three times.  In each run, a limit is imposed on the number of instances of each flow allowed to be generated during simulation.  We collect a total of six traces.  

We implement the method described in this paper in Python.  The SMT solver for solving the constraint problems in our method is the Python distribution of the Z3 solver~\cite{z3prover}.  We run our tool on the sets of collected traces, and the results are shown in Table~\ref{tab:synthetic-results}.
The first two columns show the types of and the number of unique messages in each type of traces.
The third column shows the lengths of the traces in terms of the total number of messages included. 
The fourth column shows the number of states in the extracted FSA models, while the last column shows the runtime in seconds.

\begin{table}[!t]
\renewcommand{\arraystretch}{1.3}
\caption{Results from the simulation traces}
\label{tab:synthetic-results}
\vspace{-1pt}
\centering
\begin{tabular}{ p{1.5cm}| c | c c c c }
\hline 
\bfseries Traces & \bfseries \#Messages & \bfseries Length & \bfseries \#States & \bfseries Runtime \\
\hline 
\hline
\multirow{3}{*}{\vtop{\hbox{\strut small}}} & \multirow{3}{*}{22} & $460$  & 31 & 84 \\
                                            & & $920$  & 31 & 78 \\
                                            & & $1840$  & 31 & 70 \\
\hline
\multirow{3}{*}{\vtop{\hbox{\strut large}}} & \multirow{3}{*}{60} & $2180$ & 92 & 75 \\
                                            & & $4360$ & 87 & 72 \\
                                            & & $8720$ & 100 & 62 \\
 \hline
 
\hline
\end{tabular}
\end{table} 

From Table~\ref{tab:synthetic-results}, the runtime for handling each trace takes about 1 to 1.5 minutes, and it does not change much for traces of different lengths. 
This phenomenon can be explained as follows.  
The major portion of the total runtime $(>95\%)$ is spent on the SMT solving, which is dependent on the sizes of the causality graphs in terms of the numbers of edges with non-zero support.  
The size of causality graphs depends on the number of unique messages in traces, and is much smaller than trace lengths. 
Additionally, causality graphs are relatively similar for traces of the same type. 
Therefore, solving constraint problems generated from such causality graphs incurs similar complexity,
and its runtime performance is largely independent of trace lengths.
This observation contrasts our method against previous work such as~\cite{natasa2020}. 

There are two other interesting observations from Table~\ref{tab:synthetic-results}.   
First, the runtime drops slightly for the traces of the same type; second, the runtime for the traces of different types is also similar.  These observations seem to contradict to the above analysis. 
However, we believe these results are due to the nature of SMT solving. 
For traces of the same type, they are randomly generated, resulting different constraint problems.
As a result, the initial sets of solutions generated are different,  They would lead to different amount of runtime needed to reduce every one of those initial solutions. 
Overall, the runtime taken by invoking procedures in~Algorithms~\ref{algo:model-extract} and \ref{algo:reduce-model} could be different for the traces of the same type. 
The above analysis can also explain why the runtime for traces of type {\sf Large} is close to that for {\sf small} traces. 

We compare our method with \emph{Trace2Model} described in~\cite{natasa2020}.  
It is similar to our method in that both try to synthesize FSA models to fit input traces by using some forms of constraint solving.  
However, \emph{Trace2Model} does not consider concurrency nature of the traces as our method does, and it extracts a model only based on sequential dependencies in the traces.   
In experiments, we fix the size of the sliding window $w$ in \emph{Trace2Model} to $3$ as suggested in~\cite{natasa2020}.
\emph{Trace2Model} suffers from considerably long run time when being applies to our traces, and it does complete even for the {\sf small} traces after $30$ minutes.  Therefore, no results with \emph{Trace2Model} are reported.

\begin{figure}
    \centering
    \includegraphics[height=1.5in]{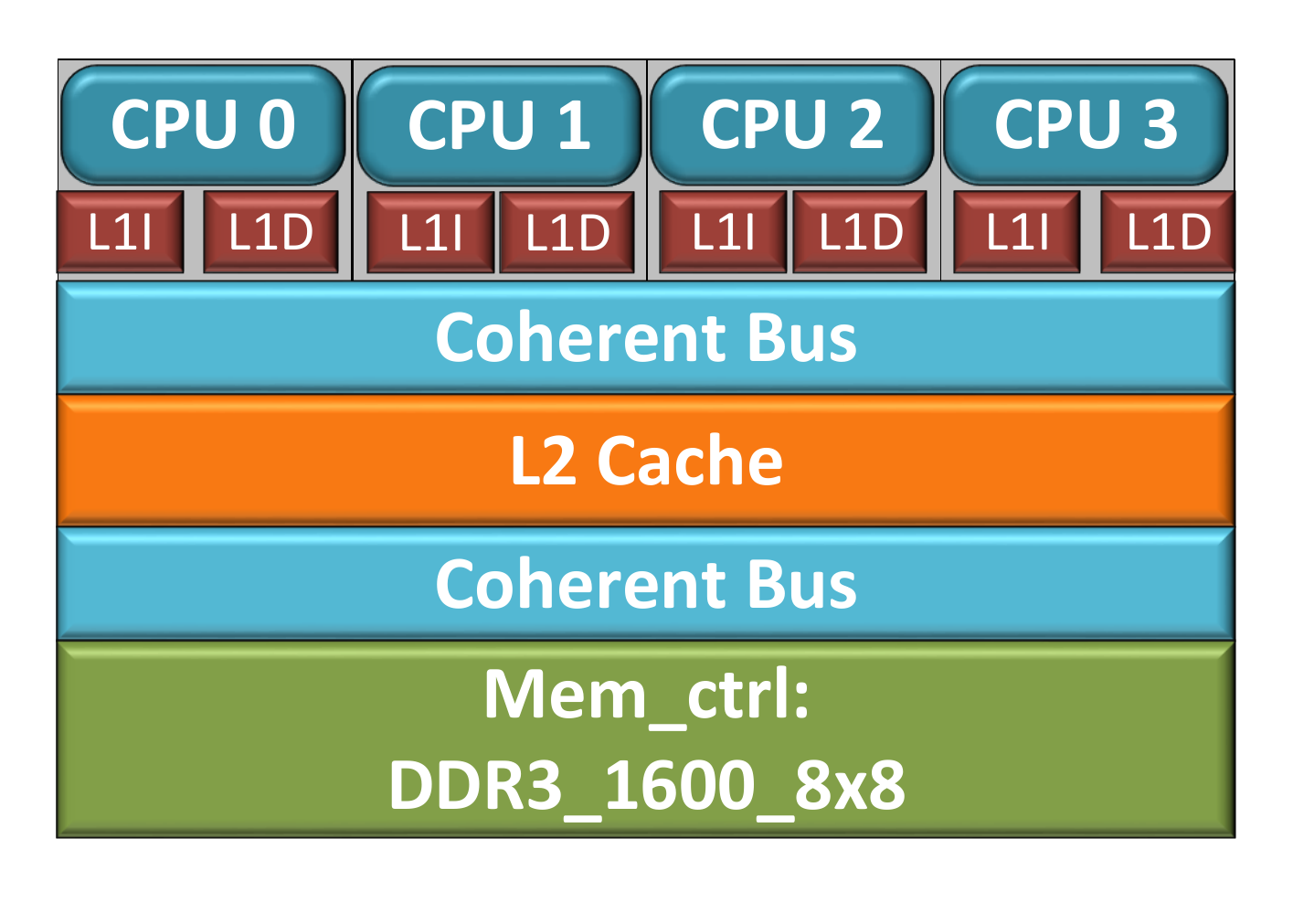}
    \caption{A quad-core simulation model developed in GEM5 SystemCall Emulation mode.}
    \label{fig:gem5-model}
\end{figure}

\begin{figure*}[tb]
\begin{tabular}{cc}
    \includegraphics[height=1.3in]{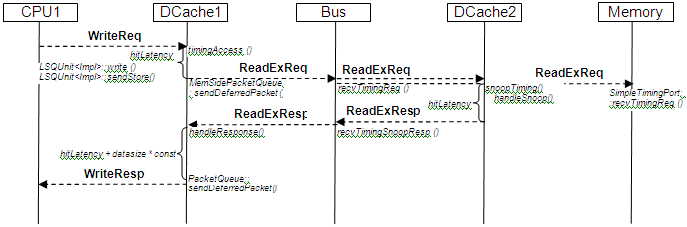}
    & 
    \includegraphics[height=1.3in]{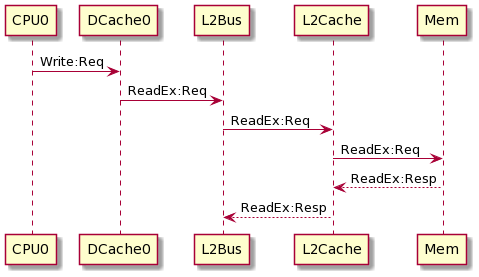}\\
    (a) & (b)
\end{tabular}
    \caption{(a) A memory write protocol specification in GEM5 documentation showing the case of write miss in DCache; (b) A path in the extracted FSA model in a sequence diagram showing the memory write operations executed by the system simulation model.}
    \label{fig:gem5-spec}
\end{figure*}

\begin{figure}[tb]
\centering
\includegraphics[height=1.2in]{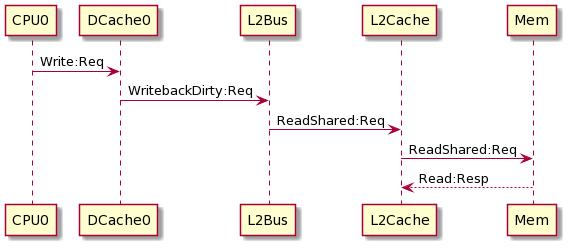}
\caption{Another example of message sequences for dirty memory block write-back operations from the extracted FSA model.}
\label{fig:mining-ex2}
\end{figure}

\medskip
\noindent{\bf Experiments on a GEM5 Model.} We develop a realistic system model in GEM5, and gather communication traces for a real-workload. 
This model, developed in Syscall Emulation~(SE) mode, consists of four x86 TimingSimpleCPUs, each with private Level 1~(L1) data~(64kB) and instruction~(16kB) caches. There is a shared Level 2~(L2) cache of size 256kB. L1 and L2 caches are interfaced with a coherent bus that implements MOESI-like coherent protocol provided in GEM5. 
There are also a memory bus and a DDR3\_1600\_8x8 memory controller with a memory of 512MB. 
A high-level view of the simulation system is given in Figure~\ref{fig:gem5-model}. 

In this GEM5 model, we insert $19$ monitors, one for each master-slave interface between two components. 
These monitors observe and dump the $packet$s (\emph{eq.} messages) exchanged over those interfaces into a trace file.  
Each packet is captured with several attributes including master/context ids, command, packet type, and address, 
We simulate this model by running a version of the Peterson's algorithm with two threads, one for each CPU core.
A trace is collected using the embedded communication monitors, and it has $785480$ packets.  
There are $103$ unique packets identified for that trace.

Using the method in this paper, we extract a model with $114$ states, which is one of models with the minimal number of states.  The process takes about $7$ minutes using $182$ MB memory at peak.  
Again, \emph{Trace2Model} in~\cite{natasa2020} is not be able to finish within $30$ minute time limit.

Figure~\ref{fig:gem5-spec}(b) shows an example of message sequences in the FSA model deserved from the trace, while Figure~\ref{fig:gem5-spec}(a) shows the similar memory write protocol from the GEM5 system documentation~\cite{gem5:protocol:write}.  
Comparing Figure~\ref{fig:gem5-spec}(b) to (a), our method is able to find the majority portion of the protocol in Figure~\ref{fig:gem5-spec}(a).  
The two messages missing in Figure~\ref{fig:gem5-spec}(b) is the result of incomplete search for solutions as described in the previous section. 
Another example of the extracted model is shown in Figure~\ref{fig:mining-ex2}.  It shows another memory write protocol where a dirty block is written back to the L2 cache under a L2 miss. 
From this example, we can find out that the GEM5 simulation model uses write-allocate policy on misses.

%% file: conclusion.tex
\section{Conclusion}
\label{sec:conclusion}

Model synthesis for communication traces from highly concurrent SoC executions is very challenging.  
We describe a method that infers reduced and abstract models from such traces, and show that it is able to infer models efficiently, even for very long traces.  The extracted models include meaningful information about system-level protocols implemented in the target SoC designs. 
We plan to enhance the method to allow user insights to be used easily to guide the search process.